\documentclass[aps,prb,twocolumn,reprint,nofootinbib,superscriptaddress,showkeys]{revtex4-2}

\usepackage{color}
\usepackage{graphicx}
\usepackage{dcolumn}
\usepackage{bm}
\usepackage{commath}
\usepackage{amsmath}
\usepackage{upgreek}
\usepackage{siunitx}
\usepackage{natbib}
\usepackage[mathlines]{lineno}
\usepackage[english]{babel}
\usepackage[latin1]{inputenc}
\usepackage[bottom=2cm,top=2cm, left=1.5cm, right=1.5cm]{geometry}
\usepackage{booktabs}
\usepackage[unicode=true, bookmarks=true, bookmarksnumbered=false, bookmarksopen=false, breaklinks=false, pdfborder={0 0 1}, backref=false, colorlinks=false]{hyperref}
\usepackage{comment}
\usepackage[normalem]{ulem}
\begin{document}

\title{Spectroscopic signatures of Kohn anomaly in a topological crystalline insulator}

\author{Jingwen Li}
\affiliation{Department of Materials, ETH Zurich, 8093 Zurich, Switzerland}

\author{Arpita Dutta}
\affiliation{School of Physical Sciences, National Institute of Science Education and Research, Jatni, 752 050 Odisha, India}
\affiliation{Homi Bhaba National Institute, Training School Complex, Anushakti Nagar, 400 094 Mumbai, India}

\author{Kush Saha}
\affiliation{School of Physical Sciences, National Institute of Science Education and Research, Jatni, 752 050 Odisha, India}
\affiliation{Homi Bhaba National Institute, Training School Complex, Anushakti Nagar, 400 094 Mumbai, India}
\affiliation{Max Planck Institute for the Physics of Complex Systems, N\"othnitzer Str. 38, 01187 Dresden, Germany}

\author{Andrzej Szczerbakow}
\altaffiliation{desceased}
\affiliation{Institute of Physics, Polish Academy of Sciences, Aleja Lotnik\'ow 32/46, 02-668 Warsaw, Poland}

\author{Tomasz Story}
\affiliation{Institute of Physics, Polish Academy of Sciences, Aleja Lotnik\'ow 32/46, 02-668 Warsaw, Poland}
\affiliation{International Research Centre MagTop, Institute of Physics, Polish Academy of Sciences, Aleja Lotnik\'ow 32/46, Warsaw 02-668, Poland}

\author{Manfred Fiebig}
\altaffiliation{manfred.fiebig@mat.ethz.ch}
\affiliation{Department of Materials, ETH Zurich, 8093 Zurich, Switzerland}

\author{Shovon Pal}
\altaffiliation{shovon.pal@niser.ac.in}
\affiliation{School of Physical Sciences, National Institute of Science Education and Research, Jatni, 752 050 Odisha, India}
\affiliation{Homi Bhaba National Institute, Training School Complex, Anushakti Nagar, 400 094 Mumbai, India}

\date{\today}

\begin{abstract}
Topological crystalline insulators extend the concept of topological insulators by hosting surface states protected by crystallographic symmetry. Their topological phase transitions arise from spin-orbit-driven band inversion in the bulk electronic structure, reshaping the low-energy electronic environment and its coupling to lattice excitations. While the electronic aspects of band topology are well established, the corresponding dynamics of lattice and electron-phonon interactions remain largely unexplored. Here, we report a pronounced softening of a low-energy surface phonon mode across the topological phase transition in Pb$_{0.77}$Sn$_{0.23}$Se, revealed by temperature-dependent time-domain terahertz spectroscopy. Unlike the well-known phonon softening in ferroelectrics, this effect does not signal a structural instability but instead reflects electronic reconstruction. We attribute the softening to the spectroscopic manifestations of a Kohn anomaly, indicating a strong coupling between lattice vibrations and Dirac-like surface electrons in the topological phase. Consistently, the phonon linewidth deviates from the standard anharmonic temperature dependence, further evidencing enhanced electron-phonon coupling. The observed changes in phonon frequency and linewidth at around 120 K are consistent with the reported topological phase transition temperature in this material. Our results thus provide a spectroscopic route for the distinct identification of the onset of a topological phase.
\end{abstract}

\maketitle

\section{Introduction}
Topological phase transitions occur when the overall topology of the electronic band structure changes. Specifically, they arise from the band inversion process that enforces the emergence of symmetry-protected gapless surface states~\cite{HasanRMP2010, BansilRMP2016}. Unlike the conventional phase transition, which involves a change in the thermodynamic Landau-type order parameter accompanied with a symmetry breaking, topological phase transitions are characterized by a change in topological invariants~\cite{YoichiJPSJ2013}. An important class of topological insulators is topological crystalline insulators (TCIs), in which the surface states are protected by the inherent crystal symmetry rather than by time-reversal symmetry~\cite{FuPRL2011}. These nontrivial surface states originate from an intertwinement of the valence and conduction bands in the Brillouin zone at specific high symmetry points, with the subsequent emergence of spin-momentum locked Dirac fermions at the surface~\cite{HsiehNA2009, XiaNP2009, RuckhoferPRR2020, GuehnePRR2024}. In contrast to conventional topological insulators, TCIs provide convenient access to topological phase transitions that can be systematically tuned by composition and temperature. 

A hallmark feature of the topological surface states is their robustness against spin-independent scattering. Due to spin-orbit coupling, the surface electrons in these states have a well-defined helicity, with the spin locked perpendicular to their momentum. Because of this spin-momentum locking, the surface states are protected from backscattering and localization, provided the perturbations they encounter are spin-independent. As a result, among the various spin-conserving scattering processes, the electron-phonon interaction becomes one of the dominant scattering channels governing the dynamics of the surface Dirac fermions at finite temperatures~\cite{ZhuPRL2012, HeidSR2017}. Notably, the emergence of symmetry-protected Dirac fermions modifies the low-energy electronic density of states at the surface~\cite{SimPRB2014, InAOM2020}. This modification directly affects the interaction between surface phonons and electronic excitations, rendering electron-phonon coupling inherently sensitive to the underlying topological phase. Thus, changes in the electron-phonon coupling strength are expected to manifest as spectroscopic signatures of topological phase transitions~\cite{ SahaPRB2014, GaratePRL2013, AntoniusPRL2016, MonserratPRL2016,ChenSC2009}. To understand the microscopic picture of these phenomena, we employ terahertz (THz) time-domain spectroscopy, where we directly investigate the low-energy electron-phonon interactions in the Pb$_{1-x}$Sn$_{x}$Se system and reveal how the coupling between surface phonons and Dirac fermions evolves across a topological phase transition.

\begin{figure*}[ht!]
    \centering
    \includegraphics[width=\textwidth]{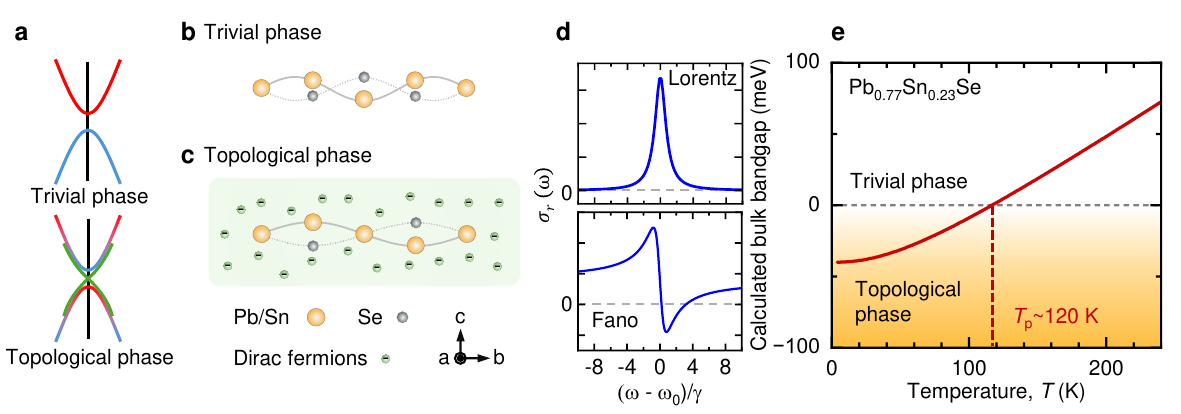}
    \caption{{\bf Trivial and topological phases in the Pb$_{1-x}$Sn$_x$Se system.} {\bf(a)} Illustration of the trivial and topological band structures of Pb$_{1-x}$Sn$_x$Se, where red denotes the bulk conduction band, blue the bulk valence band, and green the gapless surface states. {\bf(b, c)} Schematics of the shear-vertical (SV) surface phonons in the trivial and the topological phases, respectively. The green-shaded area represents the strong coupling between the Dirac fermions and the surface-localized SV phonon mode in the topological phase, which introduces a screening of the surface phonon mode, resulting in a reduction in mode frequency. Please note that the periods of the phonon oscillation are for illustration only. {\bf(d)} {\bf Top}: Schematic of the Lorentz oscillator, which describes the phonon lineshape in the THz conductivity. {\bf Bottom}: Fano asymmetry arises due to the strong electron-phonon coupling. {\bf(e)} Calculated bulk bandgap of Pb$_{0.77}$Sn$_{0.23}$Se. A positive bandgap corresponds to a normal band ordering, indicating a trivial insulating phase (white region). A negative bandgap corresponds to band inversion in the bulk, accompanied by gapless surface states, indicating a topological phase (yellow region). The red-dashed line marks the topological phase transition temperature of the sample.}
    \label{Figure1}
\end{figure*}

The Pb$_{1-x}$Sn$_{x}$Se is a well-established TCI system that offers an ideal platform for exploring electron-phonon coupling across the topological phase transition, where it changes from a topologically trivial state (with zero topological index) to an inverted band ordering with a nonzero topological index (mirror Chern number in our case), driven by an external control parameter (i.e., sample temperature), and the presence of Dirac-like surface states. The material belongs to the class of IV-VI narrow band-gap semiconductors with a rock-salt structure and hosts a wide range of electronic properties, including ferroelectricity, superconductivity, and a bulk phonon dispersion that leads to excellent thermoelectric performance~\cite{DziawaNM2012, BaronePRB2013, WoznyNJP2024, NovikovaS2018, KrizmanPRB2018}. In Pb$_{1-x}$Sn$_{x}$Se systems, the metallic surface states on (001), (110), and (111) originate from the band-inversion process at the high symmetry points in the Brillouin zone and are protected by the (1$\Bar{1}$0) mirror-plane symmetry~\cite{KalishPRL2019}. The band inversion process in Pb$_{1-x}$Sn$_{x}$Se occurs at four L points in the Brillouin zone, driven by the interplay of the relativistic effects (spin-orbit interaction and Darwin term) and the asymmetric $s-p$ orbital hybridization~\cite{WojekPRB2013, KalishPRL2019, HernandezSA2023}. Substituting Pb with Sn modifies the lattice parameters and spin-orbit coupling, leading to a closing of the bulk bandgap at a specific alloy composition. With further increase in Sn concentration, the bandgap reopens with a reversed parity of the electronic states, thereby permitting us to track the trivial-to-topological phase transition in detail~\cite{DziawaNM2012, WojekPRB2013, KalishPRL2019}. This can also be achieved by applying pressure, lowering the temperature, and engineering in-plane compressive strain in epitaxial layers~\cite{DziawaNM2012, BaronePRB2013, SzczerbakowJCG1994}. Figure~\ref{Figure1}a schematically illustrates the bulk-band dispersion in the trivial phase and the inverted band dispersion in the topological phase of Pb$_{1-x}$Sn$_{x}$Se, representing Dirac-like behavior.

In addition to its unique electronic structure, Pb$_{1-x}$Sn$_{x}$Se exhibits longitudinal optical (LO) and transverse optical (TO) phonons in the THz range, which experience strong coupling to the electronic degrees of freedom~\cite{KalishPRL2019, XiaoACSP2022}. Recent studies on the surface phonon dispersion in Pb$_{0.7}$Sn$_{0.3}$Se have unveiled the appearance of low-energy surface phonon modes, namely the shear-vertical (SV) and the shear-horizontal (SH) modes~\cite{KalishPRL2019}, arising from the overlap of the bulk optical and longitudinal acoustic phonon branches. Consequently, a pronounced reduction of the projected phonon gaps on the surface is observed near the center and the boundary of the Brillouin zone. These low-energy surface modes couple strongly with the Dirac fermions that emerge in the topological insulating phase, see Figures~\ref{Figure1}b and~\ref{Figure1}c. A characteristic signature of such electron-phonon coupling is the appearance of {\it Fano}-like lineshape in the optical conductivity, indicating quantum interference between the discrete resonance (here, the surface mode) and a broad continuum (Dirac fermions)~\cite{SimPRB2015, XuNC2017}. In the presence of a weak coupling, the phonon exhibits a symmetric resonance lineshape in the optical conductivity, while a strong interaction between electronic excitation and low-energy lattice vibration manifests an asymmetric Fano lineshape, as shown in Figure~\ref{Figure1}d. The emergence of Dirac fermions introduces a screening of the surface phonon mode, see Figure~\ref{Figure1}c, leading to a local softening or kink in the phonon dispersion -- a phenomenon referred to as the Kohn anomaly~\cite{NguyenPRL2020, ZhuPRL2011}.

\begin{figure*}
    \centering    
    \includegraphics[width=0.9\textwidth]{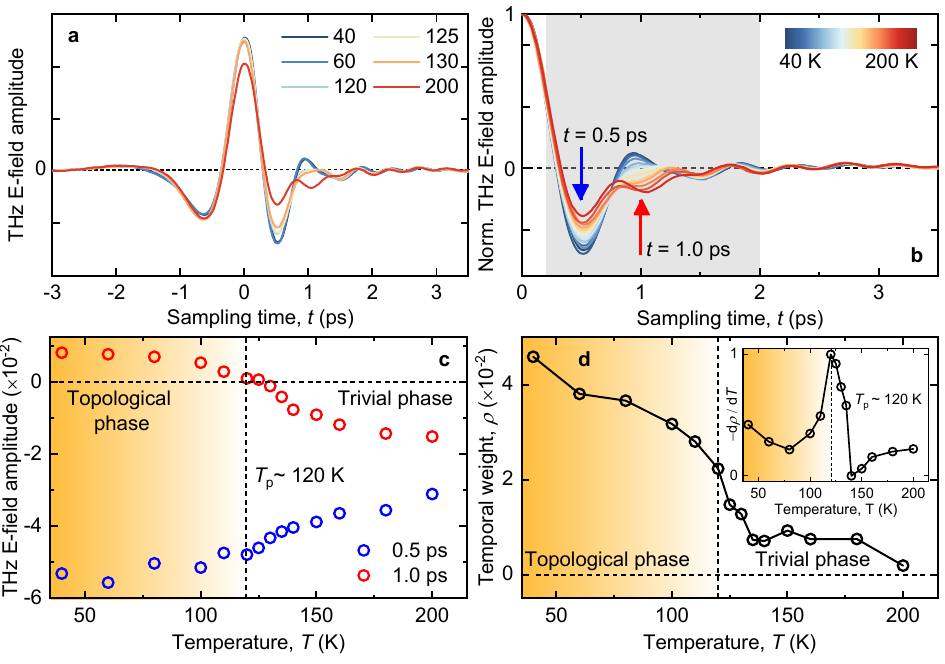}
    \caption{{\bf Reflected THz time transients.} {\bf(a)} Temperature-dependent time traces of the reflected THz electric field from Pb$_{0.77}$Sn$_{0.23}$Se for a few selected temperatures. {\bf(b)} The time window of the normalized THz signal reflected from Pb$_{0.77}$Sn$_{0.23}$Se, used to evaluate the temporal spectral weight as a function of temperature. A notable temperature dependence is observed as a post-pulse oscillation in the reflected THz signal. {\bf(c)} The amplitude of the normalized THz signal from Pb$_{0.77}$Sn$_{0.23}$Se for time stamps at $t=0.5$\,ps and 1.0\,ps. These temperature-dependent time traces highlight a notable redistribution of the THz field amplitude near the phase transition temperature, where the system evolves from a trivial to a topological phase (as indicated by the yellow-shaded region). {\bf(d)} The temperature-dependent temporal spectral weight of the time traces in the $0.2-2$\,THz range, i.e., the grey-shaded region in (b). Note the change in slope of the temporal spectral weight near the phase transition temperature. The inset shows the temperature derivative of the temporal spectral weight, which clearly identifies the transition point.}
    \label{Figure2}
\end{figure*}

In this work, using temperature-dependent THz time-domain spectroscopy, we investigate how the low-energy electrodynamics, namely, the phonon response and the electron-phonon interaction, evolve near the reported band-inversion temperature of Pb$_{0.77}$Sn$_{0.23}$Se. The reflected THz pulses, which are sensitive to this transition in both the time and frequency domains, allow us to systematically explore the electron-phonon interaction across the transition between two topologically distinct phases in Pb$_{0.77}$Sn$_{0.23}$Se. We observe a pronounced change of the Fano asymmetry in the THz conductivity of Pb$_{0.77}$Sn$_{0.23}$Se. Our data reveals a strong interaction between the Dirac fermions and the surface-localized SV phonon mode. We find that while the phonon linewidth deviates from the expected anharmonic temperature dependence, the phonon frequency displays a distinct softening, mediated by screening from Dirac fermions in the topological phase. These observations are consistent with the spectroscopic manifestation of the Kohn anomaly. In addition and quite remarkably, the temporal weight of the reflected THz transients from Pb$_{0.77}$Sn$_{0.23}$Se shows a noticeable increase with decreasing temperature, suggesting a redistribution of the low-energy spectral weight associated with the emergent surface states in the topological phase.

\begin{figure*}
    \centering
    \includegraphics[width=\textwidth]{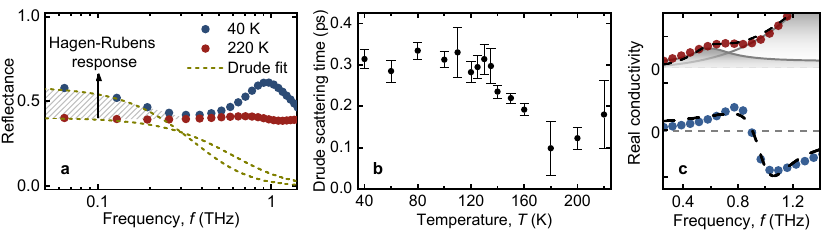}
    \caption{{\bf Reflectance and THz conductivity of Pb$_{0.77}$Sn$_{0.23}$Se.} {\bf(a)} Exemplary reflectance spectra at two different temperatures 40\,K (blue) and 220\,K (red), showing the emergence of Hagen-Rubens response (shaded-region) at lower frequencies when the system enters the topological phase. The olive-dashed lines denote the Drude-{\it only} fit of the low frequency reflectance. {\bf(b)} Temperature-dependent Drude scattering time depicting the enhanced low-temperature free carrier transport as the system evolves from the trivial to the topological phase. {\bf(c)} Exemplary spectra of real part of THz conductivity at 40\,K (blue) and 220\,K (red). The observed resonance near 1\,THz corresponds to the surface-localized shear-vertical (SV) phonon mode. As Pb$_{0.77}$Sn$_{0.23}$Se undergoes a temperature-driven trivial to topological phase transition, the lineshape of the surface phonon mode evolves from a Lorentzian-shaped two-peak spectral shape at high temperature to a pronounced Fano-like profile at low temperature, indicating the enhancement of the coupling between the SV phonon and the Dirac fermions. Note that, due to the bulk phonon around 1.8\,THz, the real conductivity increases toward higher frequency. The black dashed line represents the fit to the conductivity spectra. The grey-shaded region denotes the pure Lorentzian profiles corresponding to the surface and bulk phonons.}
    \label{Figure3}
\end{figure*}

\section{Sample and Experimental method}
The single crystals of Pb$_{1-x}$Sn$_{x}$Se samples are grown by a self-selecting vapour growth method~\cite{DziawaNM2012}. The (001)-cut sample surfaces are freshly prepared by hand polishing with Al$_2$O$_3$ polishing films with a grit size of 0.3\,$\mu$m (which is orders of magnitude smaller than $\lambda_{\rm THz}$). The sample surface orientations are verified by an X-ray single-crystal diffractometer.

Single-cycle THz pulses are generated via optical rectification in a 0.5-mm-thick (110)-cut ZnTe generation crystal, using up to 90\% of an amplified Ti:sapphire laser output (wavelength of 800\,nm, pulse duration of 120\,fs, repetition rate of 1\,kHz, pulse energy of 1.5\,mJ). The THz light pulses are then guided onto the sample using off-axis parabolic mirrors. As the Pb$_{1-x}$Sn$_{x}$Se samples are opaque to THz, we collect the signal in the 45$^\circ$-reflection geometry. We measure both the time-dependent amplitude and phase of the reflected THz light using free-space electro-optic sampling, with the residual 10\% of the laser output as the sampling beam. The THz and the sampling beams are collinearly focused onto a 0.5-mm-thick, (110)-cut ZnTe detection crystal. The THz-light-induced ellipticity of the sampling beam is then measured using a quarter-wave plate, a Wollaston prism, and a balanced photodiode. To increase the accessible sampling time between the THz and the sampling pulses, Fabry-P\'erot resonances from the faces of the detection crystal are suppressed by an additional 2-mm-thick, THz-inactive (100)-cut ZnTe single crystal, which is optically bonded to the back of the detection crystal. All measurements are performed in an inert N$_{2}$ atmosphere to avoid water absorption of the THz light. 

\section{Results and discussion}
We perform temperature-dependent THz time-domain spectroscopy~\cite{BasovRMP2011, PalPRL2019, YangPRR2020, YangNRM2023, SheePRB2024} on single crystals of Pb$_{1-x}$Sn$_{x}$Se with 23\% Sn concentration. As reported in earlier studies~\cite{DziawaNM2012, WojekPRB2014, SessiScience2016, KrizmanPRB2018}, the bandgap of Pb$_{1-x}$Sn$_{x}$Se can be tuned by either Sn concentration or temperature, leading to band inversion and a topological phase transition. Figure~\ref{Figure1}e shows the calculated temperature-dependent bandgap for Pb$_{0.77}$Sn$_{0.23}$Se illustrating the temperature-mediated topological phase transition~\cite{KrizmanPRB2018}. Note that, Pb$_{0.77}$Sn$_{0.23}$Se undergoes a topological transition at approximately 120\,K evolving from the trivial insulator to the topological insulator phase upon lowering the temperature, depicted by the red curve in Figure~\ref{Figure1}e. 

\begin{figure*}
    \centering
    \includegraphics[width=0.75\textwidth]{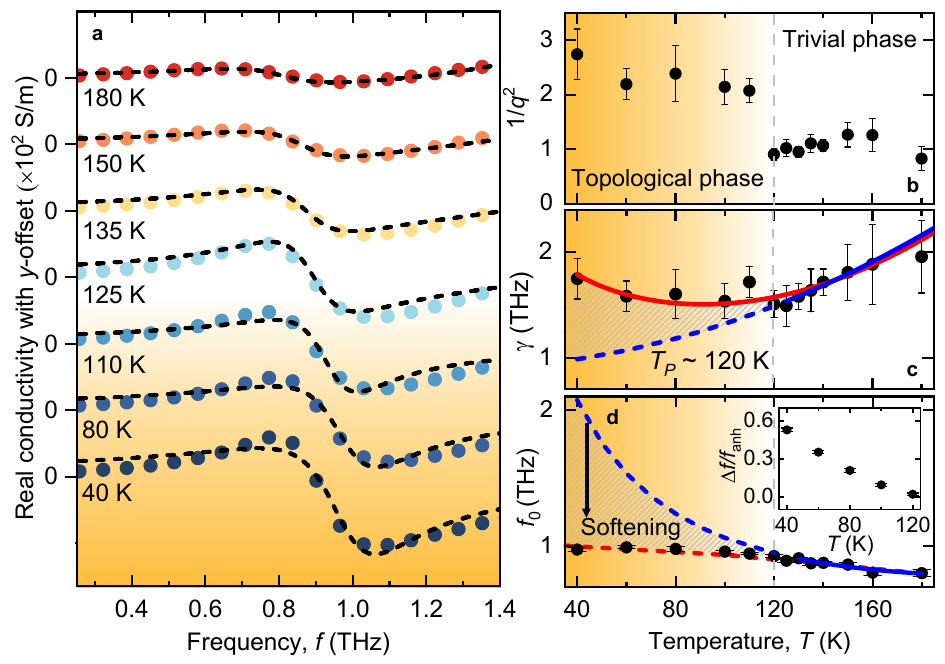}
    \caption{{\bf THz conductivity of Pb$_{0.77}$Sn$_{0.23}$Se.} {\bf(a)} Temperature-dependent real part of the THz conductivity. The solid circles are the experimental data points, while the black-dashed lines represent the Fano-model fitting using Equation~\ref{Fano}. The conductivity features a pronounced Fano asymmetry component that corresponds to the strong coupling between surface localized SV phonon and Dirac fermions in the topological phase. {\bf(b)} The temperature-dependent $1/q^2$ parameter depicts an abrupt change in the electron-phonon coupling strength near the phase transition point ($T_{\rm p}$). The phase transition (represented by the yellow shaded region) is associated with a deviation of {\bf(c)} the SV phonon linewidth from the standard anharmonicity model, along with {\bf(d)} a remarkable softening of the phonon frequency. The inset in {\bf(d)} shows the temperature dependence of the phonon softening relative to the phonon anharmonicity. In the panels, the red solid line represents the fit that includes both phonon anharmonicity and electron-phonon coupling. The red-dashed lines are guide-to-the-eye. The blue-solid lines represent the fit in accordance with the phonon anharmonicity only for $T>T_{\rm p}$, i.e., Equation~\ref{LinewidthAnh} in (c) and Equation~\ref{FrequencyAnh} in (d). The blue dashed lines represent extrapolations of the corresponding models at low temperatures. The shaded regions for $T<T_{\rm p}$ depict the linewidth deviation and the frequency softening of the SV phonon.}
    \label{Figure4}
\end{figure*}

The temperature-dependent time traces of the reflected THz signal from the Pb$_{0.77}$Sn$_{0.23}$Se are shown in Figure~\ref{Figure2}a. The time traces show an increase in the peak signal at $t=0$\,ps (the instantaneous response) with decreasing temperature. This is considered a common semiconductor behavior, in which an increase in THz reflectivity with decreasing temperature results from a combined effect of reduced lattice scattering and decreased free-carrier absorption. In the time range of $0.2-2$\,ps, the sample exhibits a prominent post-pulse oscillation that increases as temperature decreases. We quantify this buildup in two ways: by examining two distinct timestamps and by tracing the temporal weight over the time window $0.2-2$\,ps. All temperature-dependent time traces are first normalized to 1 at $t=0$\,ps, which is equivalent to scaling them to an overall identical total reflected power. 

Figure~\ref{Figure2}b shows the change in the signal amplitude at particular time stamps, i.e., $t=0.5$\,ps and $t=1$\,ps. Note that, with decreasing temperature, the modulus of oscillation amplitude at these exemplary time stamps increases as the system evolves from the trivial to the topological phase, as shown in Figure~\ref{Figure2}c. This corresponds to an overall increase in the temporal weight at the onset of the topological phase (see Figure~\ref{Figure2}d). Interestingly, the evaluated temporal weight exhibits a change in slope as the system approaches the phase-transition temperature. This behavior is clearly revealed by the negative temperature derivative (inset of Figure~\ref{Figure2}d), which is consistent with the transition temperature ($T_{\rm p}\approx120$\,K) reported earlier~\cite{DziawaNM2012, WojekPRB2014, KrizmanPRB2018}. The redistribution of the temporal weight, along with the broadening of the THz spectra (see Figure~\ref{Figure5}a of the Appendix B), captured by the reflected THz time transients at different temperatures, is attributed to the intrinsic changes in the electronic band structure. Specifically, as the system enters the topological phase, the formation of the Dirac cone enhances the density of free electrons (Dirac-like fermions), thereby modifying the low-energy excitation at the surface. 

In addition to the enhancement of the low-energy spectral weight, the sample's reflectance also exhibits a distinct temperature dependence. Two exemplary reflectance spectra (40\,K and 220\,K) are shown in Figure~\ref{Figure3}a (Refer to Figure~\ref{Figure5}b of the Appendix B for the entire set of temperature-dependent reflectance spectra). While the high-temperature reflectance remains nearly flat in the low frequency regime, a pronounced increase in reflectance is observed upon cooling, which represents the Hagen-Rubens-like response, characteristic of good conductors at low frequencies~\cite{BasovRMP2011}. We further note that at 40\,K, the Drude scattering time shows an almost three-fold increase (Figure~\ref{Figure3}b), corroborating the enhanced low-temperature free-carrier transport. The gradual change in low-energy spectral weight toward the phase transition, as well as the temperature-dependent reflectance signal, indicates the formation of Dirac-like fermions and their hybridization with surface phonon modes, owing to strong electron-phonon coupling. To directly probe the surface phonon dynamics and the underlying electron-phonon coupling, we continue to analyze the temperature-dependent THz conductivity as shown in Figure~\ref{Figure3}c. In the high-temperature conductivity spectra (i.e., 220\,K), two Lorentz-like resonances are observed, with peaks near 0.7\,THz and 1.8\,THz, corresponding to the surface-localized SV phonon mode and the bulk phonon mode, respectively. Refer to Figure~\ref{Figure6} and Appendix C for further information on the bulk phonon mode. We note that at higher temperature, the surface resonance is rather Lorentzian, which upon cooling develops an ``up-down"-like Fano-asymmetry component that becomes significantly stronger at 40\,K. This indicates an enhanced coupling between the SV phonon mode and the Dirac fermions as the system enters the topological phase below $T_{\rm p}$. In contrast, we find that the SV phonon mode is absent at all temperatures in Pb$_{0.85}$Sn$_{0.15}$Se (see Figure~\ref{Figure8} of the Appendix E), which remains in the trivial phase for the investigated temperature range.

To quantify the temperature dependence of this coupling, we model the real part of the THz conductivity using a Fano-lineshape of the form~\cite{XuNC2017}
\begin{equation}
    \sigma(\omega) = \frac{2\pi\Omega^2}{Z_0\gamma}\left[\frac{q^2 + \frac{4q(\omega-\omega_0)}{\gamma}-1}{q^2\left(1+\frac{4(\omega-\omega_0)^2}{\gamma^2}\right)}\right],
    \label{Fano}
\end{equation}
where $Z_0$ is the vacuum impedance, $\omega_0$, $\gamma$, and $\Omega$ correspond to the frequency, linewidth, and spectral strength of the surface phonon, respectively. $q$ is a dimensionless parameter that describes the asymmetry of the Fano profile. A larger $1/q^2$ represents a more pronounced asymmetry in the phonon lineshape, where the limit $1/q^2 \to 0$ recovers the symmetric Lorentzian response. The gradual evolution of the Fano-asymmetry with temperature for Pb$_{0.77}$Sn$_{0.23}$Se is shown in Figure~\ref{Figure4}a, where a dramatic increase in the $1/q^2$ parameter is observed near $T_{\rm p}$ (Figure~\ref{Figure4}b). This clearly distinguishes the two phases in Pb$_{0.77}$Sn$_{0.23}$Se, each with a distinct electronic background. The relative change of the coupling strength as the system enters the topological phase suggests a strong interaction between the low-energy electronic and phononic degrees of freedom.

To further substantiate our observations, we extract the temperature dependence of the linewidth and the frequency of the 1-THz mode as shown in Figures~\ref{Figure4}c and~\ref{Figure4}d, respectively. In the topological phase, the linewidth of the SV phonon mode deviates from the standard anharmonic behavior (as shown by the blue-dashed line in Figure~\ref{Figure4}c) and is accompanied by a pronounced softening of the resonance frequency (as shown in Figure~\ref{Figure4}d). To clarify that the softening of the phonon mode is predominantly driven by electron-phonon interactions rather than anharmonic effects, we first examine the contributions from three- and four-phonon scattering processes, which explain the decomposition of an optical phonon into two or three acoustic phonons~\cite{KlemensPR1966, BalkanskiPRB1983}. The temperature dependence of the SV phonon linewidth is captured by the equation~\cite{KlemensPR1966, SwainPRB2025}

\begin{equation}
\begin{split}
\gamma_{\,\rm anh}(T) = P_0
&+ P_1\left(1+\frac{2}{e^{y/2}-1}\right)\\
&+ P_2\left(1+\frac{3}{e^{y/3}-1}
+\frac{3}{\left(e^{y/3}-1\right)^2}\right),
\end{split}
\label{LinewidthAnh}
\end{equation}

where $y=\hbar\omega_{\,0}/k_{\rm B}T$, with $\hbar$ and $k_{\rm B}$ being the reduced Planck constant and Boltzmann constant, respectively. $P_0$, $P_1$ and $P_2$ are the anharmonic constants. Figure~\ref{Figure4}c shows that at high temperature ($T>T_{\rm p}$) the linewidth of the SV phonon mode follows the expected anharmonic model, represented by the blue solid line. The behavior, however, deviates once the system enters the topological phase (i.e., for $T<T_{\rm p}$). This deviation hints at the onset of an additional decay channel and suggests that the SV phonon mode relaxes through coupling to fermionic excitation. Such a relaxation process can arise from interband or intraband transitions involving surface Dirac fermions near the Fermi level, which become relevant at lower temperatures. This evidently alters the distribution of electronic density of states above and below the Fermi level. Considering a linearized band dispersion around the Fermi-level energy, the temperature dependence of the linewidth can be quantitatively expressed in terms of the difference in the occupation of electronic states below and above the Fermi-level energy as~\cite{XuNC2017}
\begin{equation}
    \gamma_{\,\rm e-ph}(T) = A\left[f\left(-\frac{\hbar\omega}{2k_{\rm B}T}\right)-f\left(\frac{\hbar\omega}{2k_{\rm B}T}\right)\right]
    \label{e-ph}
\end{equation}
where, A is the electron-phonon coupling constant at $T=0$\,K and $f$ is the Fermi-Dirac distribution function. To fit the temperature-dependence of the SV phonon linewidth from our THz conductivity data, we use a linewidth expression that combines both the anharmonicity and the electron-phonon coupling terms, i.e., $\gamma=\gamma_{\,\rm anh}+\gamma_{\,\rm e-ph}$. This is shown by the red-solid line in Figure~\ref{Figure4}c, which displays a remarkable agreement with our experimental data.

Figure~\ref{Figure4}d shows the temperature dependence of the SV phonon frequency at 1\,THz. For $T>T_{\rm p}$, the temperature dependence of the SV phonon frequency is captured by the equation 

\begin{equation}
\begin{split}
\omega_{\,\rm anh}(T) = \omega_{\,0}
&+ C\left(1+\frac{2}{e^{y/2}-1}\right)\\
&+ D\left(1+\frac{3}{e^{y/3}-1}
+\frac{3}{\left(e^{y/3}-1\right)^2}\right),
\end{split}
\label{FrequencyAnh}
\end{equation}

where $\omega_{\,0}$ is the bare phonon frequency in the topological phase ($\approx1$\,THz), and $C$ and $D$ are the anharmonic constants related to three-phonon and four-phonon decay processes, respectively. This is shown by the blue-solid line in Figure~\ref{Figure4}d. The blue-dashed line represents the extrapolated anharmonic nature in the topological phase ($T<T_{\rm p}$). An appreciable deviation ($\Delta f(T)>50\%$) in the SV phonon resonance frequency is observed, where $\Delta f(T) = [f_{\rm anh}(T)-f_0(T)]/f_{\rm anh}(T)$, see the inset of Figure~\ref{Figure4}d. It unambiguously signals a significant softening of the phonon mode due to screening, mediated by surface Dirac fermions --- a hallmark signature of the Kohn anomaly that Pb$_{0.77}$Sn$_{0.23}$Se displays in our experiments.

\section{Conclusion}
To conclude, we have demonstrated that THz time-domain spectroscopy provides a direct and resonant access to the low-energy lattice vibration and the electronic excitations in a Pb$_{1-x}$Sn$_{x}$Se TCI system. By systematically monitoring the temperature-mediated trivial-to-topological phase transition in Pb$_{0.77}$Sn$_{0.23}$Se, we identify clear spectroscopic signatures of electron-phonon coupling associated with the emergence of newly-formed topological surface states. The temperature evolution of a pronounced Fano asymmetry component in the real part of the THz conductivity reveals strong interference between the surface-localized SV phonon mode and the Dirac fermions as the system enters the topological phase. In addition, the deviation of the surface phonon linewidth from the standard phonon anharmonicity and the associated softening of its frequency are unambiguous spectroscopic signatures of the Kohn anomaly across the topological phase transition. Our findings suggest that electron-phonon coupling can be manipulated through composition and external tuning parameters, such as temperature and a resonant light field, thereby opening opportunities to tailor the electronic and lattice responses in topological materials. This approach is readily extendable to other topological and quantum materials, offering a pathway toward engineering a new class of optoelectronic and thermoelectric devices.

\section*{Acknowledgement}
A.D., K.S., and S.P. acknowledge the support from DAE through the projects RIN4001 and RNI4011. S.P. also acknowledges the start-up support from SERB through SERB-SRG via Project No.~SRG/2022/000290. J.L. and M.F. acknowledge the financial support from the Swiss National Science Foundation through SNSF grants Nos.~200021\_178825, 200021\_219807, and 200021\_215423. A.S. and T.S. acknowledge the partial support by the "MagTop" project (FENG.02.01-IP.05-0028/230 carried out within the "International Research Agendas" program of the Foundation for Polish Science, co-financed by the European Union under the European Funds for Smart Economy 2021-2027 (FENG).

\section*{Author Contribution}
All authors contributed to the discussion and interpretation of the experiment and to the completion of the manuscript. J.L. performed the experiments, J.L. and A.D. performed the data analysis. A.S. and T.S. provided the samples. K.S. provided the theoretical support. M.F. and S.P. conceived the project while S.P. supervised the experiments. A.D., J.L., M.F., and S.P. drafted the manuscript.

\section*{Appendix A: Phonon anharmonicity and self energy}
As discussed in the main text, Pb$_{1-x}$Sn$_x$Se is a IV-VI narrow-gap semiconductor alloy crystallizing in the rock-salt structure. Despite the high symmetry of the parent face-centered cubic lattice, lattice distortions are unavoidable due to alloying, strain, and lattice dislocations. These distortions activate various bulk phonon modes, including transverse acoustic (TA), transverse optical (TO), and longitudinal optical (LO) phonons~\cite{TianPRB2012}. Consequently, optical phonons in such systems are expected to undergo anharmonic decay through multi-phonon scattering processes. In particular, transverse optical phonons can decay into acoustic modes, contributing to a finite phonon linewidth, commonly referred to as the anharmonic contribution to the phonon linewidth. To quantify this effect, we first consider a three-phonon anharmonic process in which an optical phonon at a high-symmetry point with momentum $k=0$ decays into two acoustic phonons with momenta $k'$ and $k''$. This process satisfies both momentum conservation ($k=k'+k''$) and energy conservation ($\omega_{\,0}=\omega_{\,k'}+\omega_{\,k''}$). Treating the anharmonic interaction as a weak perturbation, the rate of change of the phonon occupation number at zero temperature ($T=0$) is given by
\begin{align}
    \frac{dN}{dt}\simeq (1+N'+N''),
\end{align}
where $N$, $N'$, $N''$ are the equilibrium distribution functions of phonon modes with momenta $k$, $k'$, $k''$, respectively. Assuming symmetric decay with $\omega_{\,k'}=\omega_{\,k''}=\omega_{\,0}/2$, the relaxation rate $\tau^{-1}$ at finite temperature ($T\neq0$) becomes
\begin{align}
    \frac{dN}{dt}=\frac{1}{\tau}
    = P\left(1+\frac{2}{e^{\beta\omega_{\,0}/2}-1}\right),
\end{align}
where $P$ is the anharmonic coupling constant associated with the three-phonon scattering process and $\beta=(k_{\rm B}T)^{-1}$. The inverse relaxation time defines the phonon linewidth, $\gamma=1/\tau$, as used in the main text. Following an analogous procedure, we next consider four-phonon scattering processes. In this case, the rate of change of the phonon occupation number is given by
\begin{align}
    \frac{dN}{dt}\simeq (1+N'+N''+N'''&+N'N''\\ \nonumber &+N'N'''+N''N'''),
\end{align}
which leads to the four-phonon contribution to the linewidth
\begin{align}
    \gamma_{\,{\rm 4-ph}}=P_1\left(1+\frac{3}{e^{\beta\omega_{\,0}/3}-1}
    +\frac{3}{(e^{\beta\omega_{\,0}/3}-1)^2}\right),
\end{align}
where $P_1$ denotes the effective anharmonic coupling constant for the four-phonon scattering process. In addition to the anharmonic effects, the strong electron-phonon coupling also contributes to the broadening of phonon linewidths. Besides bulk phonon modes, the Pb$_{1-x}$Sn$_x$Se alloys host surface phonon modes that couple strongly to electrons. This coupling gives rise to a phonon self-energy, or phonon polarization, $\Pi_\lambda(\omega,{\bf q})$, where ${\bf q}$ is the phonon wave vector, and $\omega$ is the phonon frequency for phonon modes $\lambda$. The real part of the phonon self-energy renormalizes the phonon frequency, while the imaginary part determines the phonon linewidth, which is given by $\gamma^{\,{\rm e-ph}} = -2\mathrm{Im} \Pi_\lambda({\bf q},\omega)$. Accordingly, the phonon linewidth can be expressed as
\begin{align}
    \gamma^{\,{\rm e-ph}}(T)=\frac{4\pi}{N_k}\sum_{{\bf k} n n'} &|g^{\lambda}_{n n'}({\bf k,q})|^2[f_{{\bf k}n}-f_{{\bf k-q}n'}]\\ \nonumber
    &\delta(E_{{\bf k}n}-E_{{\bf k-q}n'}-\omega_{{\bf q}\lambda}),
    \label{eq-2}
\end{align}
where the sum is on $N_k$ number of $\bf k$ vectors and $E_{{\bf k}n}$ is the energy of $n$-th band. We note that lattice vibrations in inversion-symmetric crystals can be even or odd under spatial inversion; each vibrational mode can couple to an electron. For simplicity, we assume even parity phonon modes and ${\hat g}^\lambda ({\bf q})$ to be constant. It is important to distinguish between interband ($n\neq n'$) and intraband ($n=n'$) contributions, as this distinction allows us to identify the key transitions responsible for the finite phonon linewidth observed experimentally. In a gapless spectrum, for phonons with zero momentum, only the interband contribution remains finite over the entire phonon frequency range. For finite phonon momentum, both the interband and intraband contributions may survive, depending on the momentum and temperature-dependent phonon frequencies. Assuming zone center phonon ($\bf{q}=0$) and electron-phonon coupling to be momentum independent, Equation~(\ref{eq-2}) can be recast as 
\begin{align}
    \gamma^{\,{\rm e-ph}}(T)= A (f_{{\bf k}n}(-\omega_{0\lambda}\beta/2)-f_{{\bf k}n'}(\omega_{0\lambda}\beta/2)],
\end{align}
where $A=\frac{4\pi}{N_k} |g^{\lambda}_{n n'}|^2$. We therefore employ this simplified expression in the main text to describe the phonon linewidth arising from the electron-phonon coupling.

\begin{figure}[t!]
    \centering
    \includegraphics[width=0.78\columnwidth]{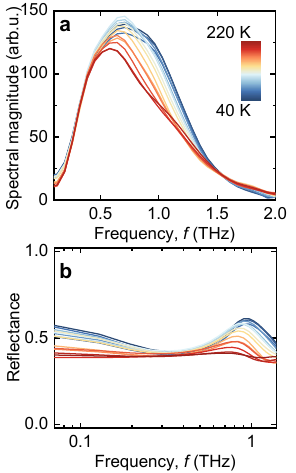}
    \caption{{\bf(a)} Temperature-dependent THz spectra and {\bf(b)} THz reflectance of Pb$_{0.77}$Sn$_{0.23}$Se. While the THz spectra show broadening of the full-width-half-maxima, the reflectance exhibits a Hagen-Rubens response in the lower-frequency region as the system approaches the topological phase at lower temperatures.}
    \label{Figure5}
\end{figure}

\begin{figure}[t!]
    \centering
    \includegraphics[width=1.0\columnwidth]{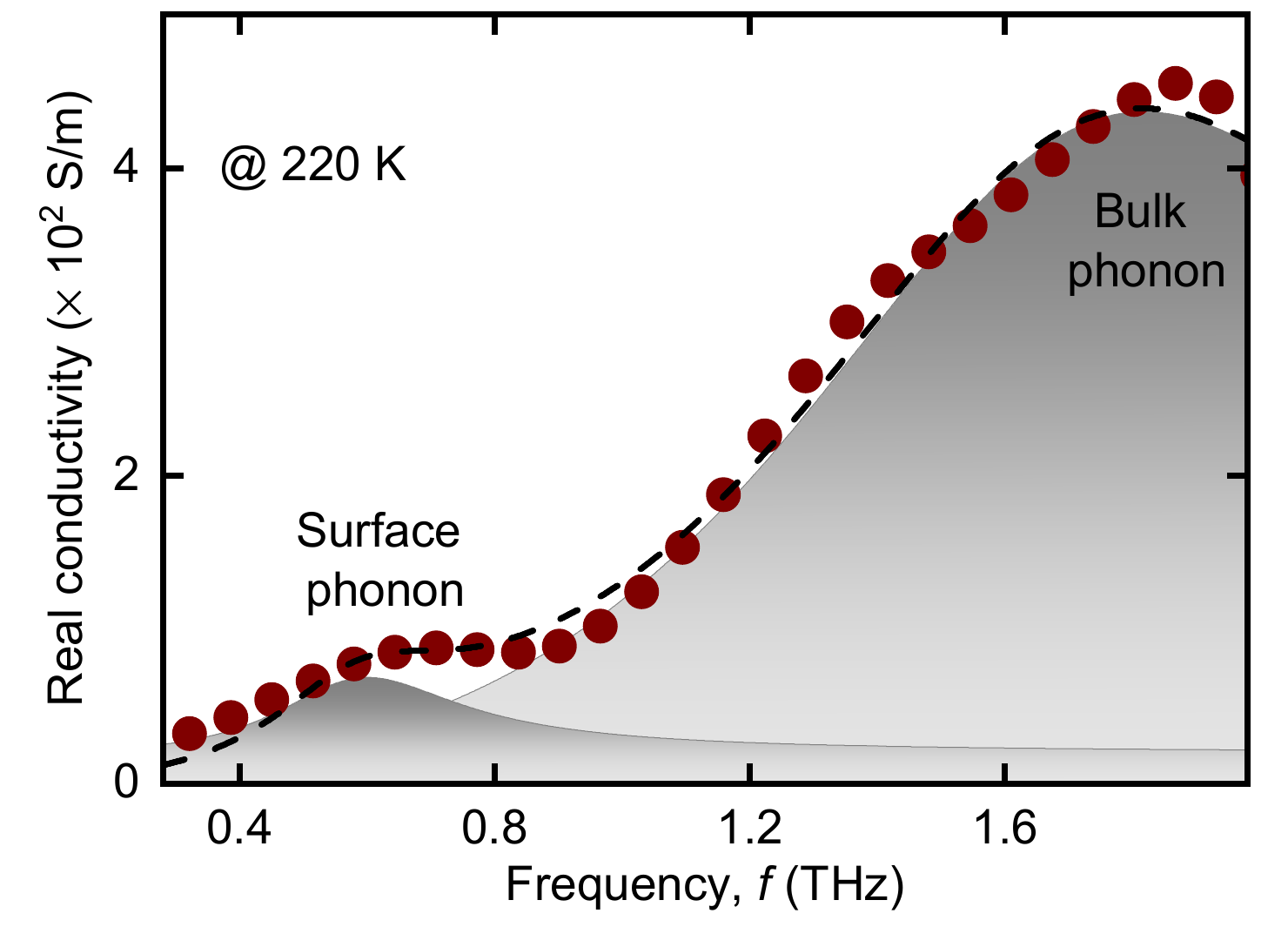}
    \vspace{-10pt}
    \caption{Real part of the THz conductivity spectra at 220\,K. The spectrum shows two pronounced resonance modes, one at 1\,THz and the other at 1.8\,THz, corresponding to the surface phonon mode and the bulk phonon mode. The black-dashed lines represent the double-Lorentz fit to the THz conductivity. The grey-shaded region denotes the pure Lorentzian profiles corresponding to the surface and bulk phonons, respectively.}
    \label{Figure6}
\end{figure}

\begin{figure}[b!]
    \centering
    \includegraphics[width=0.7\columnwidth]{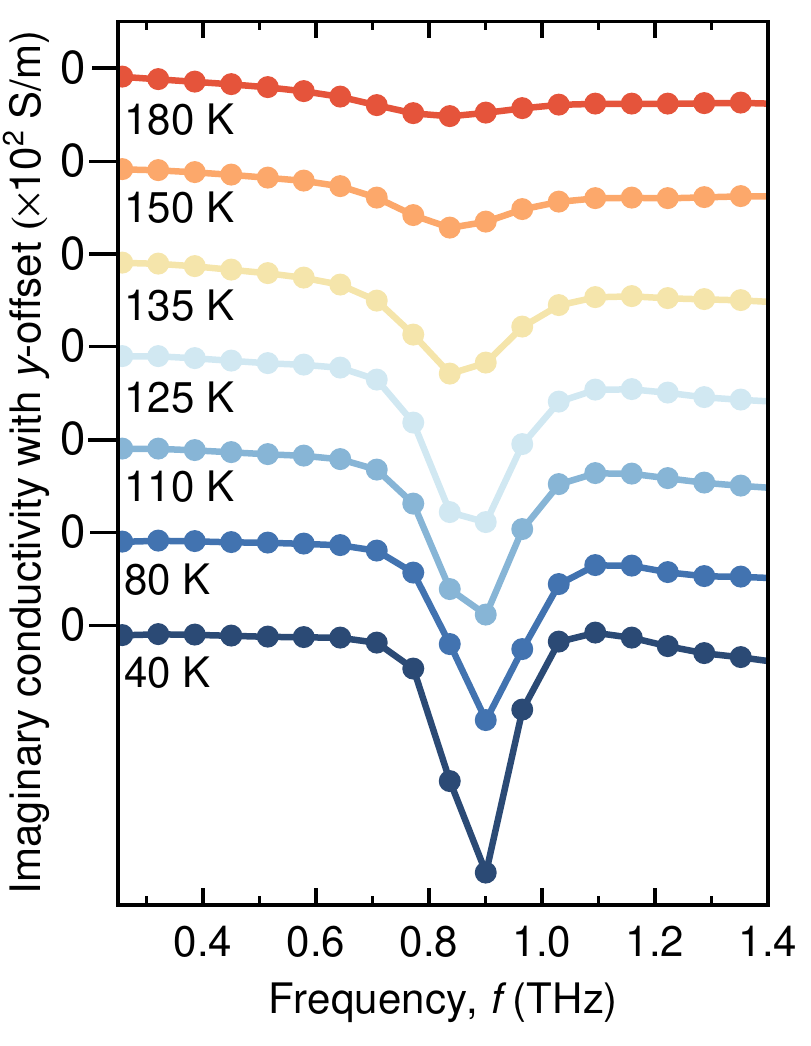}
    \caption{Imaginary part of the THz conductivity at a few selected temperatures.}
    \label{Figure7}
\end{figure}

\section*{Appendix B: Temperature-dependent THz spectra and reflectance}
The temperature-dependent THz spectra and the THz reflectance w.r.t. the high-temperature spectrum (240\,K) as reference, are plotted in Figures~\ref{Figure5}a and~\ref{Figure5}b, respectively. The FFT spectra show a broadening of the full width at half-maximum as the temperature decreases. This spectral broadening is directly linked to the change in the temporal weight slope at the onset of the topological phase transition, as discussed in detail in the main manuscript. It also corroborates the fact that an increase in spectral weight signals a redistribution of the electronic contribution in the topological phase. In addition, the THz reflectance shows a Hagen-Rubens-like response as the system evolves from a trivial to a topological phase. The enhancement of reflectance in the low-frequency regime suggests an increased carrier concentration, consistent with the emergence of newly formed Dirac-like fermions in the topological phase.

\section*{Appendix C: Bulk phonon mode}
The topological crystalline insulator sample Pb$_{0.77}$Sn$_{0.23}$Se exhibits a transverse-optical (TO) phonon mode in the THz frequency range that has been reported earlier. The frequency of this bulk phonon mode varies with the Sn concentration~\cite{NovikovaS2018}. In our system, with 23\% doping of Sn, the bulk phonon mode appeared around 1.8\,THz~\cite{XiaoACSP2022}, which is captured by the THz conductivity at 220\,K (see Figure~\ref{Figure6}). Due to the limited signal-to-noise ratio above 2\,THz, the temperature-dependent FFT spectra are reliable only up to 2\,THz. Consequently, we capture the rising edge of the broad Lorentz bulk phonon mode near 1.8\,THz. While the bulk phonon mode does not show any temperature dependence, the surface phonon near 1\,THz exhibits a pronounced temperature-dependent transition from a Lorentzian profile to a profile with a significant Fano-like component, as discussed in the main manuscript.

\section*{Appendix D: Imaginary conductivity}
The temperature-dependent imaginary part of the THz conductivity is plotted in Figure~\ref{Figure7}. Because THz time-domain spectroscopy can resolve phase information, we can also access the imaginary part of the conductivity spectrum. We note that the conductivity spectra exhibit an imaginary part of the Fano component near 1\,THz, corresponding to the resonance frequency of the surface phonon mode. With decreasing temperature, the coupling between the surface phonon and Dirac fermions becomes stronger, which is also captured in the imaginary THz conductivity, corroborating our conclusions from the real THz conductivity in Figure~4.

\section*{Appendix E: Response in trivial phase}
\begin{figure}[b!]
    \centering
    \includegraphics[width=0.8\linewidth]{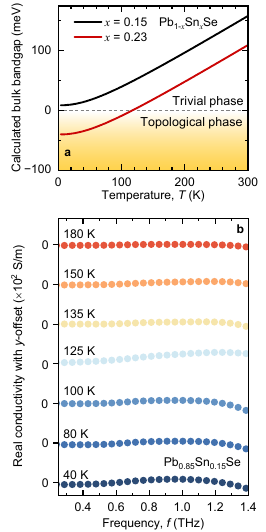}
    \caption{(a) Calculated temperature-dependent bulk band structure of Pb$_{0.85}$Sn$_{0.15}$Se, showing that the system remains in the trivial insulating phase throughout the investigated temperature range despite the reduction of the bulk band gap upon cooling. (b) Real part of the THz conductivity at selected temperatures, which shows that the 1-THz mode (as seen in Pb$_{0.77}$Sn$_{0.23}$Se) is nearly absent over the entire temperature range.}
    \label{Figure8}
\end{figure}

The sample Pb$_{0.85}$Sn$_{0.15}$Se acts as a control for the observations reported in Figure~4. The sample remains in the trivial phase throughout the investigated temperature range~\cite{DziawaNM2012} (Figure~\ref{Figure8}).

The temperature-dependent real part of the THz conductivity is shown in Figure~\ref{Figure8}b. We find that the 1-THz SV phonon mode is almost completely absent, despite the progressive reduction of the bulk band gap (Figure~\ref{Figure8}a), which is in striking contrast to what we have observed in Pb$_{0.77}$Sn$_{0.23}$Se. The markedly different behavior of these two compositions under otherwise identical experimental conditions indicates that the strong asymmetry observed in Pb$_{0.77}$Sn$_{0.23}$Se cannot be explained solely by generic carrier screening, plasmon-phonon coupling, or temperature-induced changes in the bulk electronic structure. Instead, together with the previously established identification of the 1-THz mode as the surface-localized SV phonon~\cite{KalishPRL2019} and its coupling to Dirac surface states~\cite{XiaoACSP2022}, our control measurements provide strong evidence that the observed renormalization originates from the interaction between the SV surface phonon and the emergent topological surface states.

\bibliography{Reference}

@article{HasanRMP2010,
  author  = {Hasan, M. Z. and Kane, C. L.},
  title   = {Colloquium: Topological insulators},
  journal = {Rev. Mod. Phys.},
  volume  = {82},
  pages   = {3045},
  year    = {2010},
  doi     = {10.1103/RevModPhys.82.3045}
}

@article{BansilRMP2016,
  author  = {Bansil, A. and Lin, H. and Das, T.},
  title   = {Colloquium: Topological band theory},
  journal = {Rev. Mod. Phys.},
  volume  = {88},
  pages   = {021004},
  year    = {2016},
  doi     = {10.1103/RevModPhys.88.021004}
}

@article{YoichiJPSJ2013,
  author  = {Ando, Y.},
  title   = {Topological insulator materials},
  journal = {J. Phys. Soc. Jpn.},
  volume  = {82},
  pages   = {102001},
  year    = {2013},
  doi     = {10.7566/JPSJ.82.102001}
}

@article{FuPRL2011,
  author  = {Fu, L.},
  title   = {Topological crystalline insulators},
  journal = {Phys. Rev. Lett.},
  volume  = {106},
  pages   = {106802},
  year    = {2011},
  doi     = {10.1103/PhysRevLett.106.106802}
}

@article{HsiehNA2009,
  author  = {Hsieh, D. and Xia, Y. and Qian, D. and Wray, L. A. and Dil, J. H. and Meier, F. and Fedorov, A. V. and Osterwalder, J. and Patthey, L. and Checkelsky, J. G. and Ong, N. P. and others},
  title   = {A tunable topological insulator in the spin helical Dirac transport regime},
  journal = {Nature},
  volume  = {460},
  pages   = {1101--1105},
  year    = {2009},
  doi     = {10.1038/nature08234}
}

@article{XiaNP2009,
  author  = {Xia, Y. and Qian, D. and Hsieh, D. and Wray, L. A. and Pal, A. and Lin, H. and Bansil, A. and Grauer, D. and Hor, Y. S. and Cava, R. J. and others},
  title   = {Observation of a large-gap topological-insulator class with a single Dirac cone on the surface},
  journal = {Nat. Phys.},
  volume  = {5},
  pages   = {398--402},
  year    = {2009},
  doi     = {10.1038/nphys1274}
}

@article{RuckhoferPRR2020,
  author  = {Ruckhofer, A. and Campi, D. and Bremholm, M. and Hofmann, P. and Benedek, G. and Bernasconi, M. and Ernst, W. E. and Tamt{\"o}gl, A.},
  title   = {Terahertz surface modes and electron-phonon coupling on {Bi$_2$Se$_3$(111)}},
  journal = {Phys. Rev. Research},
  volume  = {2},
  pages   = {023186},
  year    = {2020},
  doi     = {10.1103/PhysRevResearch.2.023186}
}

@article{GuehnePRR2024,
  author  = {Guehne, R. and Chlan, V.},
  title   = {Exploring the nontrivial band edge in the bulk of the topological insulators {Bi$_2$Se$_3$} and {Bi$_2$Te$_3$}},
  journal = {Phys. Rev. Research},
  volume  = {6},
  pages   = {013214},
  year    = {2024},
  doi     = {10.1103/PhysRevResearch.6.013214}
}

@article{ZhuPRL2012,
  author  = {Zhu, X. and Santos, L. and Howard, C. and Sankar, R. and Chou, F. C. and Chamon, C. and El-Batanouny, M.},
  title   = {Electron-Phonon Coupling on the Surface of the Topological Insulator {${\mathrm{Bi}}_{2}{\mathrm{Se}}_{3}$} Determined from Surface-Phonon Dispersion Measurements},
  journal = {Phys. Rev. Lett.},
  volume  = {108},
  pages   = {185501},
  year    = {2012},
  doi     = {10.1103/PhysRevLett.108.185501}
}

@article{HeidSR2017,
  author  = {Heid, R. and Sklyadneva, I. Yu. and Chulkov, E. V.},
  title   = {Electron-phonon coupling in topological surface states: The role of polar optical modes},
  journal = {Sci. Rep.},
  volume  = {7},
  pages   = {1095},
  year    = {2017},
  doi     = {10.1038/s41598-017-01180-3}
}

@article{SimPRB2014,
  author  = {Sim, S. and Brahlek, M. and Koirala, N. and Cha, S. and Oh, S. and Choi, H.},
  title   = {Ultrafast terahertz dynamics of hot Dirac-electron surface scattering in the topological insulator {${\mathrm{Bi}}_{2}{\mathrm{Se}}_{3}$}},
  journal = {Phys. Rev. B},
  volume  = {89},
  pages   = {165137},
  year    = {2014},
  doi     = {10.1103/PhysRevB.89.165137}
}

@article{InAOM2020,
  author  = {In, C. and Choi, H.},
  title   = {Dirac fermion and plasmon dynamics in graphene and 3D topological insulators},
  journal = {Adv. Opt. Mater.},
  volume  = {8},
  pages   = {1801334},
  year    = {2020},
  doi     = {10.1002/adom.201801334}
}

@article{SahaPRB2014,
  author  = {Saha, K. and Garate, I.},
  title   = {Phonon-induced topological insulation},
  journal = {Phys. Rev. B},
  volume  = {89},
  pages   = {205103},
  year    = {2014},
  doi     = {10.1103/PhysRevB.89.205103}
}

@article{GaratePRL2013,
  author  = {Garate, I.},
  title   = {Phonon-Induced Topological Transitions and Crossovers in Dirac Materials},
  journal = {Phys. Rev. Lett.},
  volume  = {110},
  pages   = {046402},
  year    = {2013},
  doi     = {10.1103/PhysRevLett.110.046402}
}

@article{AntoniusPRL2016,
  author  = {Antonius, G. and Louie, S. G.},
  title   = {Temperature-Induced Topological Phase Transitions: Promoted versus Suppressed Nontrivial Topology},
  journal = {Phys. Rev. Lett.},
  volume  = {117},
  pages   = {246401},
  year    = {2016},
  doi     = {10.1103/PhysRevLett.117.246401}
}

@article{MonserratPRL2016,
  author  = {Monserrat, B. and Vanderbilt, D.},
  title   = {Temperature Effects in the Band Structure of Topological Insulators},
  journal = {Phys. Rev. Lett.},
  volume  = {117},
  pages   = {226801},
  year    = {2016},
  doi     = {10.1103/PhysRevLett.117.226801}
}

@article{ChenSC2009,
  author  = {Chen, Y. L. and Analytis, J. G. and Chu, J.-H. and Liu, Z. K. and Mo, S.-K. and Qi, X.-L. and Zhang, H. J. and Lu, D. H. and Dai, X. and Fang, Z. and others},
  title   = {Experimental realization of a three-dimensional topological insulator, {${\mathrm{Bi}}_{2}{\mathrm{Te}}_{3}$}},
  journal = {Science},
  volume  = {325},
  pages   = {178--181},
  year    = {2009},
  doi     = {10.1126/science.1173034}
}

@article{DziawaNM2012,
  author  = {Dziawa, P. and Kowalski, B. J. and Dybko, K. and Buczko, R. and Szczerbakow, A. and Szot, M. and {\L}usakowska, E. and Balasubramanian, T. and Wojek, B. M. and Berntsen, M. H. and others},
  title   = {Topological crystalline insulator states in {${\mathrm{Pb}}_{1-x}{\mathrm{Sn}}_{x}{\mathrm{Se}}$}},
  journal = {Nat. Mater.},
  volume  = {11},
  pages   = {1023--1027},
  year    = {2012},
  doi     = {10.1038/nmat3449}
}

@article{WojekPRB2014,
  author  = {Wojek, B. M. and Dziawa, P. and Kowalski, B. J. and Szczerbakow, A. and Black-Schaffer, A. M. and Berntsen, M. H. and Balasubramanian, T. and Story, T. and Tjernberg, O.},
  title   = {Band inversion and the topological phase transition in {(Pb,Sn)Se}},
  journal = {Phys. Rev. B},
  volume  = {90},
  pages   = {161202(R)},
  year    = {2014},
  doi     = {10.1103/PhysRevB.90.161202},
   addednote = {NEW REFERENCE}
}

@article{SessiScience2016,
  author  = {Sessi, P. and Di Sante, D. and Szczerbakow, A. and Glott, F. and Wilfert, S. and Schmidt, H. and Bathon, T. and Dziawa, P. and Greiter, M. and Neupert, T. and others},
  title   = {Robust spin-polarized midgap states at step edges of topological crystalline insulators},
  journal = {Science},
  volume  = {354},
  pages   = {1269--1273},
  year    = {2016},
  doi     = {10.1126/science.aah5064}
}

@article{BaronePRB2013,
  author  = {Barone, P. and Rauch, T. and Di Sante, D. and Henk, J. and Mertig, I. and Picozzi, S.},
  title   = {Pressure-induced topological phase transitions in rocksalt chalcogenides},
  journal = {Phys. Rev. B},
  volume  = {88},
  pages   = {045207},
  year    = {2013},
  doi     = {10.1103/PhysRevB.88.045207}
}

@article{SzczerbakowJCG1994,
  author  = {Szczerbakow, A. and Berger, H.},
  title   = {Investigation of the composition of vapour-grown {${\mathrm{Pb}}_{1-x}{\mathrm{Sn}}_{x}{\mathrm{Se}}$} crystals {($x\leq 0.4$)} by means of lattice parameter measurements},
  journal = {J. Cryst. Growth},
  volume  = {139},
  pages   = {172--178},
  year    = {1994},
  doi     = {10.1016/0022-0248(94)90048-5}
}

@article{WoznyNJP2024,
  author  = {Wo{\'z}ny, M. and Szuszkiewicz, W. and Dyksik, M. and Motyka, M. and Szczerbakow, A. and Bardyszewski, W. and Story, T. and Cebulski, J.},
  title   = {Electron--phonon coupling and a resonant-like optical observation of a band inversion in topological crystal insulator {Pb$_{1-x}$Sn$_x$Se}},
  journal = {New J. Phys.},
  volume  = {26},
  pages   = {063008},
  year    = {2024},
  doi     = {10.1088/1367-2630/ad4e67}
}

@article{NovikovaS2018,
  author  = {Novikova, N. N. and Yakovlev, V. A. and Kucherenko, I. V. and Vinogradov, V. S. and Aleschenko, Y. A. and Muratov, A. V. and Karczewski, G. and Chusnutdinow, S.},
  title   = {Infrared Reflection Spectra of {Pb$_{1-x}$Sn$_x$Se ($x=0.2$, $0.34$)} Topological Insulator Films on a ZnTe/GaAs Substrate and the Vibrational Modes of Multilayer Structures},
  journal = {Semiconductors},
  volume  = {52},
  pages   = {34--40},
  year    = {2018},
  doi     = {10.1134/S1063782618010191}
}

@article{KrizmanPRB2018,
  author  = {Krizman, G. and Assaf, B. A. and Phuphachong, T. and Bauer, G. and Springholz, G. and de Vaulchier, L. A. and Guldner, Y.},
  title   = {Dirac parameters and topological phase diagram of {${\mathrm{Pb}}_{1-x}{\mathrm{Sn}}_{x}{\mathrm{Se}}$ from magnetospectroscopy}},
  journal = {Phys. Rev. B},
  volume  = {98},
  pages   = {245202},
  year    = {2018},
  doi     = {10.1103/PhysRevB.98.245202}
}

@article{KalishPRL2019,
  author  = {Kalish, S. and Chamon, C. and El-Batanouny, M. and Santos, L. H. and Sankar, R. and Chou, F. C.},
  title   = {Contrasting the Surface Phonon Dispersion of {${\mathrm{Pb}}_{0.7}{\mathrm{Sn}}_{0.3}{\mathrm{Se}}$} in Its Topologically Trivial and Nontrivial Phases},
  journal = {Phys. Rev. Lett.},
  volume  = {122},
  pages   = {116101},
  year    = {2019},
  doi     = {10.1103/PhysRevLett.122.116101}
}

@article{HernandezSA2023,
  author  = {Hernandez, F. G. G. and Baydin, A. and Chaudhary, S. and Tay, F. and Katayama, I. and Takeda, J. and Nojiri, H. and Okazaki, A. K. and Rappl, P. H. O. and Abramof, E. and others},
  title   = {Observation of interplay between phonon chirality and electronic band topology},
  journal = {Sci. Adv.},
  volume  = {9},
  pages   = {eadj4074},
  year    = {2023},
  doi     = {10.1126/sciadv.adj4074}
}

@article{XiaoACSP2022,
  author  = {Xiao, Z. and Wang, J. and Liu, X. and Assaf, B. A. and Burghoff, D.},
  title   = {Optical-pump terahertz-probe spectroscopy of the topological crystalline insulator {Pb$_{1-x}$Sn$_x$Se} through the topological phase transition},
  journal = {ACS Photonics},
  volume  = {9},
  pages   = {765--771},
  year    = {2022},
  doi     = {10.1021/acsphotonics.1c01717}
}

@article{SimPRB2015,
  author  = {Sim, S. and Koirala, N. and Brahlek, M. and Sung, J. H. and Park, J. and Cha, S. and Jo, M.-H. and Oh, S. and Choi, H.},
  title   = {Tunable Fano quantum-interference dynamics using a topological phase transition in {${(\mathrm{Bi}_{1-x}\mathrm{In}_{x})}_{2}\mathrm{Se}_{3}$}},
  journal = {Phys. Rev. B},
  volume  = {91},
  pages   = {235438},
  year    = {2015},
  doi     = {10.1103/PhysRevB.91.235438}
}

@article{XuNC2017,
  author  = {Xu, B. and Dai, Y. M. and Zhao, L. X. and Wang, K. and Yang, R. and Zhang, W. and Liu, J. Y. and Xiao, H. and Chen, G. F. and Trugman, S. A. and others},
  title   = {Temperature-tunable Fano resonance induced by strong coupling between Weyl fermions and phonons in {TaAs}},
  journal = {Nat. Commun.},
  volume  = {8},
  pages   = {14933},
  year    = {2017},
  doi     = {10.1038/ncomms14933}
}

@article{NguyenPRL2020,
  author  = {Nguyen, T. and Han, F. and Andrejevic, N. and Pablo-Pedro, R. and Apte, A. and Tsurimaki, Y. and Ding, Z. and Zhang, K. and Alatas, A. and Alp, E. E. and Chi, S. and Fernandez-Baca, J. and Matsuda, M. and Tennant, D. A. and Zhao, Y. and Xu, Z. and Lynn, J. W. and Huang, S. and Li, M.},
  title   = {Topological Singularity Induced Chiral Kohn Anomaly in a Weyl Semimetal},
  journal = {Phys. Rev. Lett.},
  volume  = {124},
  pages   = {236401},
  year    = {2020},
  doi     = {10.1103/PhysRevLett.124.236401}
}

@article{ZhuPRL2011,
  author  = {Zhu, X. and Santos, L. and Sankar, R. and Chikara, S. and Howard, C. and Chou, F. C. and Chamon, C. and El-Batanouny, M.},
  title   = {Interaction of Phonons and Dirac Fermions on the Surface of {${\mathrm{Bi}}_{2}{\mathrm{Se}}_{3}$}: A Strong Kohn Anomaly},
  journal = {Phys. Rev. Lett.},
  volume  = {107},
  pages   = {186102},
  year    = {2011},
  doi     = {10.1103/PhysRevLett.107.186102}
}

@article{BasovRMP2011,
  author  = {Basov, D. N. and Averitt, R. D. and van der Marel, D. and Dressel, M. and Haule, K.},
  title   = {Electrodynamics of correlated electron materials},
  journal = {Rev. Mod. Phys.},
  volume  = {83},
  pages   = {471--541},
  year    = {2011},
  doi     = {10.1103/RevModPhys.83.471}
}

@article{PalPRL2019,
  author  = {Pal, S. and Wetli, C. and Zamani, F. and Stockert, O. and L{\"o}hneysen, H. v. and Fiebig, M. and Kroha, J.},
  title   = {Fermi Volume Evolution and Crystal-Field Excitations in Heavy-Fermion Compounds Probed by Time-Domain Terahertz Spectroscopy},
  journal = {Phys. Rev. Lett.},
  volume  = {122},
  pages   = {096401},
  year    = {2019},
  doi     = {10.1103/PhysRevLett.122.096401}
}

@article{YangPRR2020,
  author  = {Yang, C.-J. and Pal, S. and Zamani, F. and Kliemt, K. and Krellner, C. and Stockert, O. and L{\"o}hneysen, H. v. and Kroha, J. and Fiebig, M.},
  title   = {Terahertz conductivity of heavy-fermion systems from time-resolved spectroscopy},
  journal = {Phys. Rev. Res.},
  volume  = {2},
  pages   = {033296},
  year    = {2020},
  doi     = {10.1103/PhysRevResearch.2.033296}
}

@article{TianPRB2012,
  author  = {Tian, Z. and Garg, J. and Esfarjani, K. and Shiga, T. and Shiomi, J. and Chen, G.},
  title   = {Phonon conduction in {PbSe, PbTe, and PbTe$_{1-x}$Se$_x$} from first-principles calculations},
  journal = {Phys. Rev. B},
  volume  = {85},
  pages   = {184303},
  year    = {2012},
  doi     = {10.1103/PhysRevB.85.184303}
}

@article{YangNRM2023,
  author  = {Yang, C.-J. and Li, J. and Fiebig, M. and Pal, S.},
  title   = {Terahertz control of many-body dynamics in quantum materials},
  journal = {Nat. Rev. Mater.},
  volume  = {8},
  pages   = {518--532},
  year    = {2023},
  doi     = {10.1038/s41578-023-00555-7}
}

@article{SheePRB2024,
  author  = {Shee, P. and Yang, C.-J. and Kumar Pandey, S. and Kumar Nandy, A. and Kulkarni, R. and Thamizhavel, A. and Fiebig, M. and Pal, S.},
  title   = {Terahertz crystal electric field transitions in a Kondo-lattice antiferromagnet},
  journal = {Phys. Rev. B},
  volume  = {109},
  pages   = {075133},
  year    = {2024},
  doi     = {10.1103/PhysRevB.109.075133}
}

@article{KlemensPR1966,
  author  = {Klemens, P. G.},
  title   = {Anharmonic decay of optical phonons},
  journal = {Phys. Rev.},
  volume  = {148},
  pages   = {845},
  year    = {1966},
  doi     = {10.1103/PhysRev.148.845}
}

@article{BalkanskiPRB1983,
  author  = {Balkanski, M. and Wallis, R. F. and Haro, E.},
  title   = {Anharmonic effects in light scattering due to optical phonons in silicon},
  journal = {Phys. Rev. B},
  volume  = {28},
  pages   = {1928--1934},
  year    = {1983},
  doi     = {10.1103/PhysRevB.28.1928}
}

@article{SwainPRB2025,
  author  = {Swain, D. and Dey, A. and Roy, A. and Saha, K. and Das, S. D.},
  title   = {Nontrivial phonon dynamics and significant electron-phonon coupling of the high-frequency modes in a Dirac semimetal},
  journal = {Phys. Rev. B},
  volume  = {111},
  pages   = {035143},
  year    = {2025},
  doi     = {10.1103/PhysRevB.111.035143}
}

@article{WojekPRB2013,
  author  = {Wojek, B. M. and Buczko, R. and Safaei, S. and Dziawa, P. and Kowalski, B. J. and Berntsen, M. H. and Balasubramanian, T. and Leandersson, M. and Szczerbakow, A. and Kacman, P. and Story, T. and Tjernberg, O.},
  title   = {Spin-polarized (001) surface states of the topological crystalline insulator {Pb$_{0.73}$Sn$_{0.27}$Se}},
  journal = {Phys. Rev. B},
  volume  = {87},
  pages   = {115106},
  year    = {2013},
  doi     = {10.1103/PhysRevB.87.115106}
}
\end{document}